\documentstyle[12pt] {article}

\newcommand{\beq}{\begin{equation}}
\newcommand{\eeq}{\end{equation}}
\newcommand{\beqa}{\begin{eqnarray}}
\newcommand{\eeqa}{\end{eqnarray}}

\begin{document}

\begin{center}
{\large \bf SWKB for the Angular Momentum}
\vskip 1. truecm

{\bf Luca Salasnich}\footnote{E--mail: salasnich@math.unipd.it} 
\vskip 0.3cm
Dipartimento di Matematica Pura ed Applicata, Universit\`a di Padova, \\
Via Belzoni 7, 35131 Padova, Italy \\
Istituto Nazionale di Fisica Nucleare, Sezione di Padova, \\
Via Marzolo 8, 35131 Padova, Italy \\
Istituto Nazionale di Fisica della Materia, Unit\`a di Milano, \\
Via Celoria 16, 20133 Milano, Italy \\

\vskip 0.5 truecm

{\bf Fabio Sattin}\footnote{
{\it Present address}: 
Istituto Gas Ionizzati del C.N.R.,
Corso Stati Uniti 4, 35127 Padova, Italy. E--mail: sattin@igi.pd.cnr.it} 
\vskip 0.3cm
Dipartimento di Ingegneria Elettrica, Universit\`a di Padova, \\
Via Gradenigo 6/a, 35131 Padova, Italy  \\
Istituto Nazionale di Fisica della Materia, Unit\`a di Padova, \\
Corso Stati Uniti 4, 35127 Padova, Italy 

\end{center}

\newpage

\begin{center}
{\bf Abstract} 
\end{center}
\vskip 0.5 truecm
\par
It has been recently shown [M. Robnik and L. Salasnich, 
{\it J. Phys. A: Math. Gen.}, {\bf 30}, 1719 (1997)] that 
the WKB series for the quantization of angular momentum $L$ converges 
to the exact value $L^2=\hbar^2 l(l+1)$, if summed over all orders, and 
gives the Langer formula $L^2=\hbar^2 (l+1/2)^2$ at the leading order. 
In this work we solve the eigenvalue problem 
of the angular momentum operator by using the supersymmetric 
semiclassical quantum mechanics (SWKB), and show that
it gives the correct quantization already at the leading order. 

\vskip 0.5cm

\begin{center}
\end{center}

\newpage

\par 
The Wentzel--Kramers--Brillouin (WKB) semiclassical method 
is one of the most useful techniques for solving the Schr\"odinger 
equation. It allows to obtain approximate 
analytic expressions for the wavefunctions and energy spectra${}^{1)}$ 
Usually, what is used is the torus quantization, which is just 
the leading order of the $\hbar$-expansion of the WKB method. 
Higher terms can be calculated with a recursion formula in one degree systems, 
but are generally unknown in systems with more than one degree of freedom 
(see Robnik and Salasnich${}^{2)}$). 
\par
It has been observed since a long time that the WKB method, in its leading
approximation, when applied to three--dimensional spherically symmetric 
problems yields wrong results, unless one replaces the correct value
$L^2 = \hbar^2 l (l+1)$ with the Langer expression${}^{3)}$ 
$L^2 = \hbar^2 (l+1/2)^2$. Further, when higher order terms 
are included, the Langer correction needs modifications at each order of
approximation. It has recently been shown by Robnik and Salasnich${}^{2)}$
that the Langer term and its first corrections are the first terms
of an infinite series. By making a guess on the higher order terms, 
they could sum all the series, recovering the exact quantum result. 
\par
In the framework of the supersymmetric 
semiclassical quantization (SWKB), Comtet, Bandrauk and Campbell${}^{4)}$ 
obtained at the leading order the exact quantization of the radial 
part of the Kepler problem by using the correct value 
$L^2 = \hbar^2 l (l + 1)$. 
The aim of the present work is to complete their result. In fact, we show that 
also the exact quantization of the angular momentum is obtained 
at the first order of the SWKB expansion. 
\par
The eigenvalue equation of the angular momentum operator is 
\beq
\hat{L}^2 Y(\theta , \phi) = \lambda^2 \hbar^2 Y(\theta , \phi) \; ,
\eeq
with
\beq
\hat{L}^2 = \hat{P}_\theta^2 + { \hat{P}_{\phi}^2 \over \sin^2(\theta)} 
          = - \hbar^2 \left( { \partial^2 \over \partial \theta^2 } +
\cot(\theta) {\partial \over \partial \theta} \right) - \hbar^2
{ 1\over \sin^2(\theta)}  {\partial^2 \over \partial \phi^2} \; . 
\eeq
After the substitution
\beq
Y(\theta , \phi ) = T(\theta) e^{ i m \phi} \; ,
\eeq
we obtain the equation 
\beq
T''(\theta) + \cot(\theta) T'(\theta) + \left( \lambda^2 - 
{ m^2 \over \sin^2(\theta) } \right) T(\theta) = 0 \; ,
\eeq
where $m$ is the azimuthal quantum number.  
This equation is exactly solvable. Its eigenvalues 
and eigenfunctions are well known from 
any text of quantum mechanics (see, {\it e.g.} Landau and Lifshitz${}^{5)}$):
the former are $\lambda^2=l(l+1)$, $l \geq m$; the latter are the 
associate Legendre polynomials. 
\par
Now we will briefly outline the WKB expansion of Eq. (4). First of all,
it is important to notice that in (4) does not appear $\hbar$, 
therefore an expansion in powers of this parameter is not possible. 
To override this difficulty a small parameter $\epsilon$ 
is introduced${}^{6)}$:
\beq
\epsilon^2 T''(\theta) + \epsilon^2 \cot(\theta) T'(\theta) + 
\left( \lambda^2 - { m^2 \over \sin^2(\theta) } \right) T(\theta) = 0 \; .
\eeq
The parameter $\epsilon$, which will be set to 1 at the end of the calculation,
has formally the same role of $\hbar$ as ordering parameter, and is equivalent
to taking the limit $\lambda^2 \to \infty$ and/or $ m \to \infty$. 
With the further substitution
\beq
T(\theta) = \exp \left( {1 \over \epsilon} {\sum_{n \geq 0} S_n \epsilon^n} 
\right) \; ,
\eeq
we find
\beq
(S_0')^2 +  \left( \lambda^2 - { m^2 \over \sin^2(\theta) }  \right) = 0 \; ,
\eeq
\beq
{\sum_{k=0}^n S_k' S_{n-k}'} + S_{n-1}'' + \cot(\theta) S_{n-1}' = 0 
\; ,  \;\;\;\; n > 0  \; .
\eeq
The exact quantization is obtained by requiring the uniqueness 
of the wave function: 
\beq
\oint dS = \sum_{k=0}^{\infty} \oint dS_k = 2 \pi i \; n_{\theta} \; ,
\eeq
where we have now set $\epsilon =1$. This integral is a complex contour 
integral which encircles the two turning points on the real axis. 
Obviously, it is derived from the requirement of the uniqueness of the 
complex wave function $T$. If one stops to the leading order 
\beq 
\oint dS_0 = \oint i \sqrt{  \left( \lambda^2 -
{ m^2 \over \sin^2(\theta) } \right)} \; d \theta = 
2 \pi i \left( n_{\theta} + { 1\over 2} \right) \; .
\eeq
The integral is easily calculated and the condition 
\beq
\lambda = n_{\theta} + m + {1\over 2} = l + {1\over 2} 
\eeq
is obtained. Otherwise, one may go further, compute all the 
terms $S_n$ and recover the exact result $\lambda^2 = l (l + 1)$,  
as done by Robnik and Salasnich${}^{6)}$. 
\par
After these preliminaries, we enter into the main topic of this paper:
we shall repeat the calculation using the supersymmetric theory 
(for a monography about the subject of SUSY quantum mechanics see, {\it e.g.}, 
Junker${}^{7)}$). 
\par
In order to put (4) in a standard form we make the replacement
\beq
T(\theta) = { F(\theta) \over \sqrt{ \sin(\theta)} } \; ,
\eeq
and obtain
\beq
F''(\theta) + \left[ \left( \lambda^2 + {1 \over 4} \right) +
{ 1 \over \sin^2(\theta) } \left( { 1\over 4 } - m^2 \right)
\right] F(\theta) = 0 \; .
\eeq
This equation has the standard form of the one--dimensional Schr\"odinger 
equation with $\hbar=2M=1$. Its eigenvalues are $(\lambda^2 + 1/4)$. 
\par
To perform the SWKB for the Eq. (4) or (13), 
it is necessary to calculate the ground state 
wave--function $T_0(\theta )=\sin^m{(\theta )}$ 
and its eigenvalue $\lambda_0=m(m+1)$. 
Then we can define the supersymmetric (SUSY) potential
\beq
\Phi(\theta) = - { d \ln ( F_0 (\theta)) \over d \theta} =
 - \left(  m + { 1 \over 2} \right) \cot(\theta) \; ,
\eeq
with
\beq
F_0(\theta) = T_0(\theta) \sqrt{\sin(\theta)} \; .
\eeq 
From $\Phi$ the two SUSY partner potentials and Hamiltonians
may be defined
\beq
H_{\pm} = - {d^2 \over d \theta^2} + V_{\pm}(\theta) \; ,
\eeq
\beq
V_{\pm}(\theta) = \Phi^2(\theta)  \pm \Phi'(\theta) \; .
\eeq
The following statements hold : i) the ground-state energy of $H_{-}$,
$E_{-}^0$, vanishes; ii) all other eigenvalues of $H_{-}$, $E_{-}$, coincide 
with that of $H_{+}$; iii) the spectrum of $H_{-}$ and that of (12)
differ by a constant:
\beq
E_{-} = \lambda^2 + { 1\over 4} - \left( \lambda_0^2 + { 1\over 4}
\right) \; , 
\eeq
where $\lambda_0=m(m+1)$ is the eigenvalue 
of the ground state of Eq. (4). We do 
not give here the proofs since they are standard results of the SUSY theory 
(see Ref. 7 and references therein for details).
\par
A further important remark is that $\Phi$ gives rise to {\it shape--invariant} 
partner potentials: $V_{-}$ and $V_{+}$ depend, besides the independent
variable $\theta$, from the parameter $m$. A couple of $V_{-} , V_{+}$,
depending upon an independent variable $x$ and a set of parameters $\{a_0 \}$,
are called shape--invariant if the relation holds:
\beq 
V_{-}(a_0 , x) = V_{+}(a_1 , x) + R(a_1) \; ,
\eeq
with $ \{a_1 \}$ a new set of parameters and $R$ a function of $a_1$ but not of
$x$. In our case we obtain 
\beq
V_{-} (m , \theta) = V_{+}(m-1 ,\theta) - 2 m \; ,
\eeq
It has been shown by Dutt, Khare and Sukhatme${}^{8)}$, and by 
Barclay and Maxwell${}^{9)}$, that for this class of potentials:
i) exact quantization rules are attained at the first order, and the results
may be expressed in analytical form; 
ii) higher order terms are identically null.
\par 
Let us apply the WKB formalism to $H_{-}$ of Eq. (16); the main lines of
the calculation follow Eqns. (5--10), but a careful analysis 
(see Ref. 7, ch. 6) shows that 
the quantization conditions are to be modified: to the leading order one
gets
\beq
{\int_a^b \sqrt{ E_{-} - \Phi^2(x)} \; d x } = n_{\theta} \pi
\eeq
with $a , b$ roots of 
\beq
E_{-} - \Phi^2(x) = 0
\eeq
This formula is also referred to as CBC formula, from 
Comtet, Bandrauk and Campbell${}^{4)}$. Notice that: i) on the left hand
side there appears $\Phi^2$ instead of the full 
potential $V_{-}$; ii) in the right hand side 
the Maslov correction (the 1/2 term) does not appear any longer.
\par
Substituting for $\Phi$ in (21), (22) the expression (14), one easily finds
\beq
{\int_a^b \sqrt{ E_{-} - \Phi^2(x)} \; d x } =
 \pi \left[ \sqrt{ E_{-} + \left( m + {1 \over 2} \right)^2 }
- \left( m + {1 \over 2} \right) \right] \\
= n_{\theta} \pi \; ,
\eeq
with $b = - a = \arctan \sqrt{ \left( m + { 1 \over 2} \right) 
\over E_{-} }$. In conclusion, we have
\beq
E_{-} = \left( n_{\theta} + m + {1 \over 2} \right)^2 -
\left( m + {1 \over 2} \right)^2 \; . 
\eeq
Now, by using Eq. (18) with $\lambda_0 =m (m+1)$, we get  
\beq
\lambda^2 = (n_{\theta} + m) ( n_{\theta} + m + 1) \; ,
\eeq
which, with the position $l = n_{\theta} + m$, yields the result sought.
\par
Summarizing, we have demonstrated that, by using SUSY quantum
mechanics, the eigenvalue problem of the angular momentum operator can 
be solved exactly within the semiclassical approximation and 
at the lowest order. Even if these results are a consequence of 
the theorems of SUSY theory, 
we think that an explicit derivation is not devoid of interest 
since, notwithstanding the importance of the three--dimensional 
spherically symmetric problems in physics, 
a detailed analysis of its angular part has always been lacking in literature
until the work of Robnik and Salasnich${}^{6)}$. That work, however, suffered
from the fact that its results were based upon some conjectures which,
though fully reasonable, were not rigorously demonstrated. 
\vskip 0.5cm
\begin{center}
*****
\end{center}
\par
FS has been supported during this work by a grant of the Italian MURST.

\newpage

\section*{References}

\parindent = 0.pt

1. V.P. Maslov and M.V. Fedoriuk, {\it Semi-Classical Approximations in 
Quantum Mechanics} (Reidel Publishing Company, 1981). 

2. M. Robnik and L. Salasnich, {\it J. Phys. A: Math. Gen.} {\bf 30},
   1711 (1997).

3. R.I. Langer, {\it Phys. Rev.}, {\bf 51}, 669 (1937).

4. A. Comtet, A.D. Bandrauk, and D.K. Campbell, {\it Phys. Lett.},
{\bf B 150}, 159 (1985).

5. L.D. Landau and E.M. Lifshitz, {\it Quantum Mechanics} (3rd Edition,
Pergamon, 1977).

6. M. Robnik and L. Salasnich, {\it J. Phys. A: Math. Gen.} {\bf 30},
   1719 (1997).

7. G. Junker, {\it Supersymmetric Methods in Quantum and Statistical
 Physics} (Springer, 1996).

8. R. Dutt, A. Khare, and U.P. Sukhatme, {\it Phys. Lett.}, {\bf B 181}, 295
   (1986).
 
9. D.T. Barclay and C.J. Maxwell, {\it Phys. Lett.}, {\bf A 157}, 357 (1991). 

\end{document}